# The Tunka-Rex Experiment for the Detection of the Air-Shower Radio Emission

[a]Y. Kazarina[1], P.A. Bezyazeekov[1], N.M. Budnev[1], O.A. Gress[1], A. Haungs[2], R.Hiller[2], T. Huege[2], M. Kleifges[3], E.N. Konstantinov[1], E.E. Korosteleva[4], D.Kostunin[2], O. Krömer[3], L.A. Kuzmichev[4], R.R. Mirgazov[1], L. Pankov[1], V.V.Prosin[4], G.I. Rubtsov[5], C. Rühle[3], V. Savinov[1], F.G. Schröder[2], R.Wischnewski[6], A. Zagorodnikov[1] (Tunka-Rex Collaboration)

[1]*Institute of Applied Physics ISU, Irkutsk, Russia*
[2]*Institut für Kernphysik, Karlsruhe Institute of Technology (KIT), Germany*
[3]*Institut für Prozessdatenverarbeitung und Elektronik, KIT, Germany*
[4]*Skobeltsyn Institute of Nuclear Physics MSU, Moscow, Russia*
[5]*Institute for Nuclear Research of the Russian Academy of Sciences, Moscow, Russia*
[6]*DESY, Zeuthen, Germany*

[a]Corresponding author: lutien777@mail.ru

**Abstract.** The Tunka-Rex experiment (Tunka Radio Extension) has been deployed in 2012 at the Tunka Valley (Republic of Buryatia, Russia). Its purpose is to investigate methods for the energy spectrum and the mass composition of high-energy cosmic rays based on the radio emission of air showers. Tunka-Rex is an array of 25 radio antennas distributed over an area of 3 km$^2$. The most important feature of the Tunka-Rex is that the air-shower radio emission is measured in coincidence with the Tunka-133 installation, which detects the Cherenkov radiation generated by the same atmospheric showers. Joint measurements of the radio emission and the Cherenkov light provide a unique opportunity for cross calibration of both calorimetric detection methods. The main goal of Tunka-Rex is to determine the precision for the reconstruction of air-shower parameters using the radio detection technique. In this article we present the current status of Tunka-Rex and first results, including reconstruction methods for parameters of the primary cosmic rays.

## 1  INTRODUCTION

One of the main questions of astroparticle-physics is the explanation of the observed features of the spectrum of ultra-high-energy cosmic rays (UHECR) and their assignment to galactic or extragalactic origin. "Classic" primary cosmic rays are atomic nuclei accelerated to high energies up to $E_0 \geq 10^{20}$[1]. The acceleration of cosmic rays to such high energies remains an unsolved mystery. High-energy cosmic rays are studied by registering products of their interaction with the atmosphere. As a result of this interaction extensive air showers (EAS) develops, consisting of muonic, nuclear, and electromagnetic cascades. Almost all elementary particles may be present in EAS at high energies, but mainly electrons, muons, γ-rays, neutrinos, Cherenkov radiation, fluorescence radiation and radio emission reach the Earth's surface [2].

To answer the open questions high statistics as well as excellent measurement quality for the UHECR is needed, which requires the application of new detection technologies.

The Tunka-Rex experiment (Tunka Radio Extension) is deployed at the Tunka Valley at the site of the Tunka-133 Cherenkov array [3] (figure 1). The Tunka-133 experiment measures Cherenkov light emitted by air showers. It is located close to Irkutsk, Siberia, Russia, and consists of 175 photomultiplier detectors in a hexagonal structure.

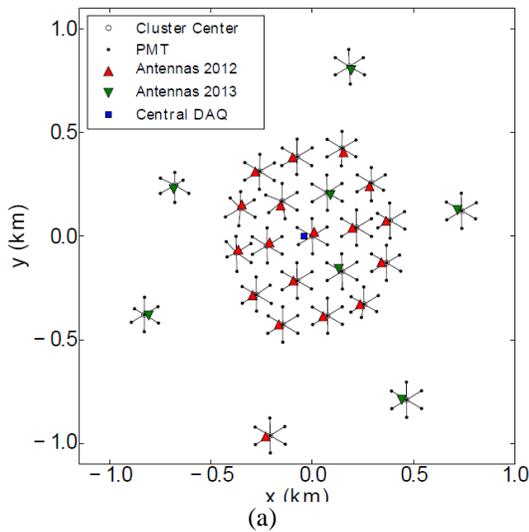 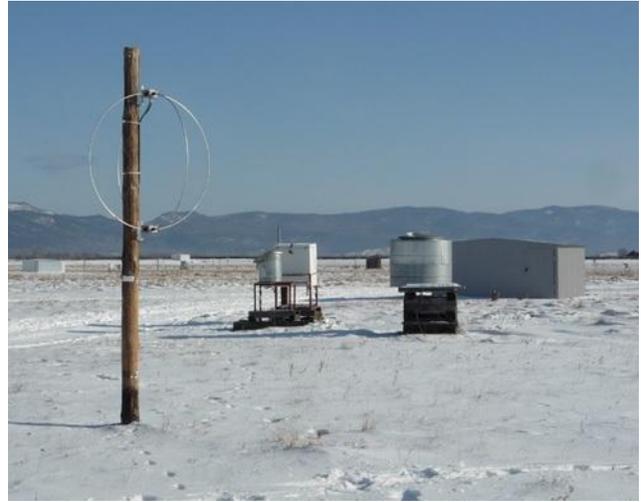

(a) (b)

**FIGURE 1**. (a) Scheme of Tunka-Rex: antennas are placed at the centers of the Tunka-133 clusters (= hexagons of 7 photomultiplier detectors). (b) Tunka-Rex antenna station, left, together with a Tunka-133 and a HiSCORE stations.

Since Cherenkov measurements are sensitive to the longitudinal shower development, they feature a relatively good estimation of the primary energy and mass. The main advantages of the Cherenkov array are high accuracy for the EAS core position (5-10 m), good angular resolution ~ 0.1 – 0.3˚, a good energy resolution (~ 15%), and the capability to statistically estimate the mass of the primary particle via the atmospheric depths of the shower maximum $X_{max}$ (precision 25 g/cm$^2$) [4]. But the main disadvantage of Tunka-133 is the short time of operation: the duty cycle is only 5-10% because the detection of the Cherenkov light requires moonless, cloudless nights.

Nowadays Tunka-133 is being extended. The new TAIGA array (Tunka Advanced Instrument for cosmic ray physics and Gamma Astronomy) will be a complex hybrid detector for ground-based gamma-ray astronomy for energies from a few TeV to several PeV as well as for cosmic ray studies from 100 TeV to several EeV. TAIGA will include Tunka-HiSCORE - an array of wide-angle integrating air-Cherenkov stations, an array of imaging atmospheric Cherenkov telescopes, and an array of particle detectors, both on the surface and underground [5].

## 1.1 Tunka-Rex: Hardware and Calibration

Tunka-Rex measurements are carried out jointly with the installation Tunka-133, which records the Cherenkov radiation generated by the same air showers. Joint measurements of the radio emission and the Cherenkov light provide a unique opportunity for cross-calibration of the two calorimetric methods for air-shower detection. The power of the coherent radio signal increases quadratically with the shower energy, which leads to a detection threshold of approximately 0.1 EeV for Tunka-Rex. Advantages of EAS radio detection are the relative cheapness and simplicity of the radio antennas and the high duty cycle: the radio method is independent of the time of day and of weather conditions (except thunderstorms) – in contrast to the optical detection methods, which need clear and moonless nights. The optimal frequency range for registration of the coherent radio emission emitted by EAS is 30-80 MHz. For coherent radiation, the wavelength must be larger than the thickness of the shower disk (~ m). In principle, on could measure also at lower frequencies, but this is difficult, since both the natural and anthropogenic background increases towards lower frequencies.

Tunka-Rex currently consists of 25 antennas attached by cables to the local data-acquisition of the clusters of the Tunka-133 photomultiplier array, which is organized in 25 clusters formed by 7 PMTs each. The spacing between the antennas in the inner clusters is approximately 200 m, covering an area of roughly 1 km$^2$. The total area including the outer clusters is approximately 3 km$^2$. At each antenna position there are two orthogonally aligned SALLAs [6] with 120 cm diameter (figure 1). Unlike in most radio experiments [7, 8], and like in the LOFAR experiment [9], the antennas in Tunka-Rex are not aligned along the north-south and east-west axis, but rotated by 45˚. Since the radio signal from cosmic ray air showers is predominantly east-west polarized, this should result in more antennas with signal in both channels but also less events with signal in at least one channel.

Although the SALLA provides a signal-to-noise ratio worse compared to other antenna types, it has two major advantages. First, it is economic, compact and simple. Second, the gain of the SALLA is relatively independent from environmental conditions. This is important, since the measurement accuracy of pulses with high amplitude is not limited by noise, but by variations of the gain due to environmental changes and ground conditions.

The signal chain is continued by a low noise amplifier (LNA) placed in a metal box in the top of the SALLA. A 30 meter long, buried coaxial cable connects the antenna to the main amplifier and a filter at each cluster center of Tunka-133 which hosts a flash ADC board for the digital data acquisition. Each of the clusters features its own local DAQ which is triggered by the PMTs. There, the signal from both, the antennas and the PMTs, is digitized and transmitted to the central DAQ via optical fibers where it is stored on disk.

To reconstruct the real, physical signal from the measured signal it is necessary to calibrate all parts of the signal chain. The antenna calibration was made at the Karlsruhe Institute of Technology, Germany, with the same calibration source used already in LOPES [10]. This is a calibrated reference source with known electric field, which generates peaks with a period of 1 microsecond. In the frequency domain this is a 1 MHz frequency comb. The calibration source was put above the antenna and the received signal was recorded with an oscilloscope. Then, the scale of the absolute amplitude is derived from the ratio between the emitted and the received power.

Thus, using this method we calibrate almost the whole signal chain at once, except of the last missing part in the signal chain namely the ADC, since for technical reason the calibration measurement is not performed with the ADC used for the real measurements, but with an oscilloscope. First we checked the linearity of the ADC with a peak signal generator and generated a calibration curve. Surprisingly it deviates from the specifications in the data sheet (figure 2). Thus, we also measured the frequency behavior of the ADC with a sine signal generator. Interestingly, the response varies up to 20 % over frequency which is now taken into account in the signal reconstruction of cosmic-ray measurements.

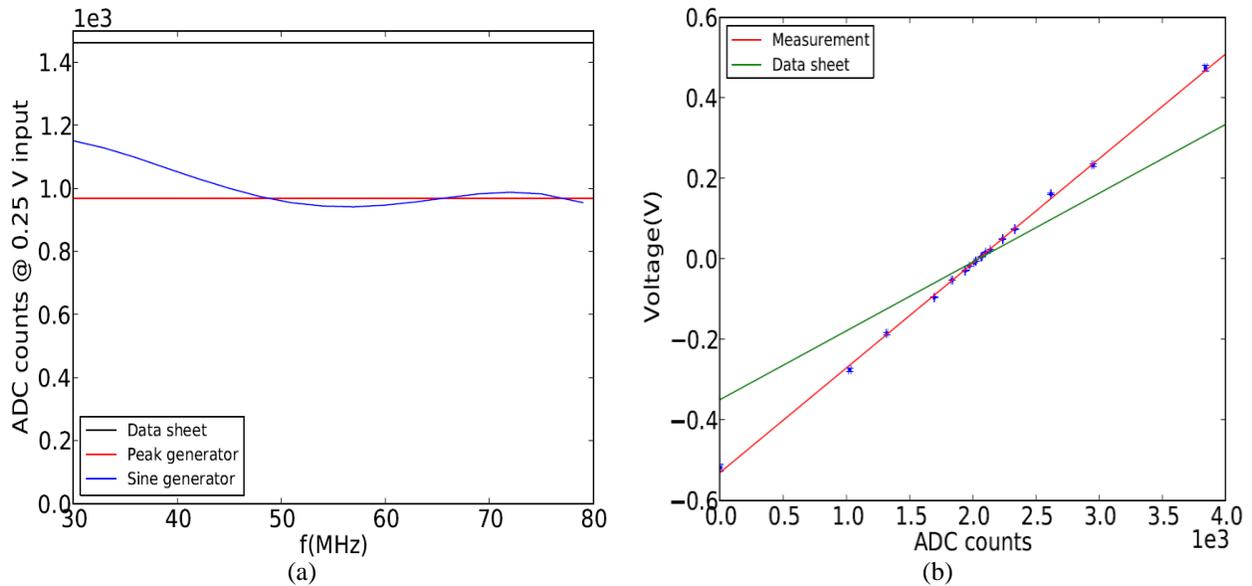

**FIGURE 2**. ADC calibration. (a) Amplitude frequency characteristic of the ADC. (b) Linearity of the ADC.

## 2    DATA ANALYSIS

Until now we had 2 seasons of data taking, namely Oct 2012 – Apr 2013, and Oct 2013 – Apr 2014. The Tunka-Rex collaboration has decided to perform a semi-blind analysis for our main goal, i.e., the determination of the Tunka-Rex precision for the reconstruction of the energy and the shower maximum by comparing to the air-Cherenkov measurements. That means, we develop and tune methods for the radio reconstruction using the first season of measurements (2012/2013) and blind the air-Cherenkov reconstruction for the second season. For the second season, the energy and $X_{max}$ values are kept secret by the Tunka-133 collaboration, and will be revealed to the Tunka-Rex collaboration later. Now, only the direction and core position of the Tunka-133 reconstruction are available to cross-check which radio pulses are emitted by air-showers, and which are background pulses.

The total measurement time of the first season is approximately 400 hours [11]. In this period we identified 62 air-shower events with a zenith angle $\leq 50°$, and 84 events with larger zenith angles. All events had to fulfill the following quality cuts:
- angle between shower axes reconstructed by Cherenkov and radio detectors $< 5°$;
- at least 3 antennas with a signal-to-noise ratio (SNR) $> 6$ at the trigger time.

To test the sensitivity of the Tunka-Rex measurements to the air shower parameters we reconstructed the lateral distribution of the radio signal for the events of the first season with zenith angles $\leq 50°$ (see figure 3 for an example).

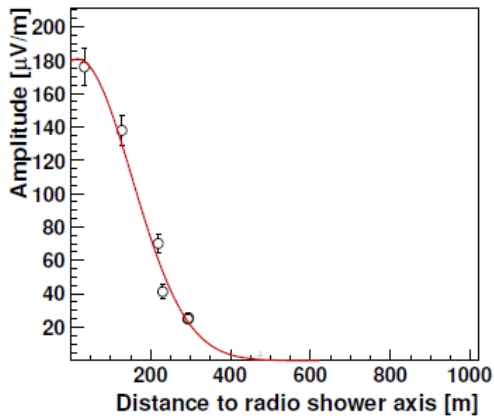

**FIGURE 3**. Lateral distribution of one example event with a Gaussian fit function.

It is known that the amplitude parameter of the lateral distribution function (LDF) is correlated with the primary energy [12]. Moreover, the slope of the lateral distribution is sensitive to the position of the shower maximum [12, 13], which will be analyzed in near future by comparing Tunka-Rex measurements to the $X_{max}$ reconstruction of the PMT array Tunka-133. For the energy reconstruction, we observe a correlation between the radio field strength at 100 m corrected for the geomagnetic effect and the energy reconstructed with the air-Cherenkov measurements (figure 4).

The given values for the field strength are based on a Gaussian fit function. The main conclusion is that apart from a few outliers, the average deviation between the radio and the air-Cherenkov reconstruction is in the order of the air-Cherenkov energy precision: 15 %, only. The outliers are under investigation. Also when using an exponential lateral distribution function instead, we observe a similar correlation for the amplitude at 100 m with the primary energy. Thus, for the energy reconstruction, the exponential LDF is a sufficient approximation. It offers the advantage of a larger statistics, because it only has to degrees of freedom and, consequently, 3 antennas are sufficient for the exponential LDF.

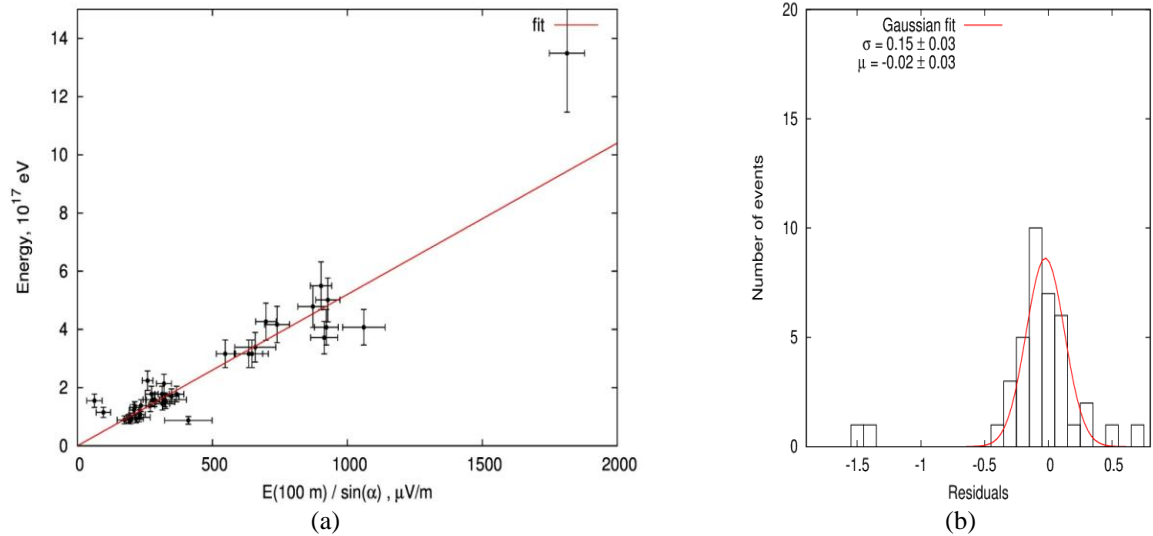

**FIGURE 4.** (a) Correlation between the energy reconstructed with the air-Cherenkov measurements and the radio field strength at 100 m corrected for the geomagnetic effect (we divide the radio field strength at 100 m by sin α, where α is a geomagnetic angle, i.e. the angle between the shower axis and the geomagnetic field). (b) The residuals in the right hand plot are calculated as follows: $(E_{cherenkov} - E_{radio})/E_{cherenkov}$

# 3 CONCLUSION

Results of the 2012/2013 season have shown that Tunka-Rex successfully measured the radio emission of air showers with energies above $10^{17}$ eV in combination with the Tunka air-Cherenkov array. Thus, Tunka provides hybrid measurements which enable a cross-calibration between the air-Cherenkov and the radio signal. After quality cuts there is strong correlation between radio amplitude and energy, reconstructed by the air-Cherenkov array. This year we plan to connect new Tunka-Rex antennas to the scintillator extension of Tunka. This enables day-time measurements and consequently increases the duty cycle. In addition, we plan to equip some of the HiSCORE [14] detector stations in the Tunka array with additional antennas. The additional antennas will increase the efficiency of the detector and possibly also the accuracy for the reconstruction of the energy and the mass composition. After we have finally developed our methods for the reconstruction of $X_{max}$, we will unblind the data of the season 2013/2014 to cross-check the results, and to determine the reconstruction precision of Tunka-Rex in this semi-blind analysis.

# ACKNOWLEDGMENTS

We acknowledge the support of the Russian Federation Ministry of Education and Science (G/C 14.518.11.7046, 14.B25.31.0010, 14.14.B37.21.0785, 14.B37.21.1294), the Russian Foundation for Basic Research (Grants 11-02-00409, 13-02 00214, 13-02-12095, 13-02-10001), the VPO ISU (Grant 111-13-203) and by the Helmholtz Association (Grant HRJRG-303) and the Helmholtz Alliance for Astroparticle physics (HAP). We thank the Pierre Auger Collaboration for their permission to use the *Offline* analysis software [15] for the Tunka-Rex analysis.

# REFERENCES

[1]   V.A. Tsarev. Ultrahigh energy cosmic rays registration by radio method. Physics of Elementary Particles and Nuclei, V. 35, pp. 186-247, 2004.
[2]   N.M. Budnev et al., Experimental studies of high and ultra-high-energy primary cosmic rays on the "TUNKA". The lectures of Baikal School of Fundamental Physics, pp. 3-8, 2005.

[3]  V. Prosin, et al. (Tunka-133 Collaboration), Nucl. Instr. Meth. A 756 (2014) 94
[4]  S.F. Berezhnev et al. (Tunka-133 Collaboration), Proc. 32$^{nd}$ ICRC Beijing, China (2011) # 0495.
[5]  N.M. Budnev et al., TAIGA the Tunka Advanced Instrument for cosmic ray physics and Gamma Astronomy — present status and perspectives. Journal of Instrumentation, V. 9, 2014
[6]  The Pierre Auger Collaboration, JINST 7 (2012) P10011.
[7]  H. Falcke, et al. (LOPES Collaboration), Nature 435 (2005) 313.
[8]  The Pierre Auger Collaboration, Phys. Rev. D 89 (2014) 052002.
[9]  P. Schellart, et al. (LOFAR Collaboration), Astronomy and Astrophysics 560 (2013) A98.
[10] S. Nehls, et al., Nucl. Instr. Meth. A 589 (2008) 350.
[11] D. Kostunin, et al. (Tunka-Rex Collaboration), Nucl. Instr. Meth. A 742 (2014) 89 (Proc. 13rd RICAP).
[12] W.D. Apel, et al. (LOPES Collaboration), Phys. Rev. D 90 (2014) 062001.
[13] W.D. Apel, et al. (LOPES Collaboration), Phys. Rev. D 85 (2012) 071101(R).
[14] M. Tluczykont, et al, Astroparticle Physics 56 (2014) 42.
[15] The Pierre Auger Collaboration, Nucl. Instr. Meth. A 635 (2011) 92.